\documentclass[a4paper]{article}
\usepackage[latin1]{inputenc} 
\usepackage[T1]{fontenc} 
\usepackage{RR,RRthemes}
\usepackage{graphicx}
\usepackage{amsmath}
\usepackage{algorithm}
\usepackage{subfigure}

\newcommand{\ones}{\underline{1}}

\newtheorem{proposition}{Proposition}

\RRdate{October 2011}

\RRauthor{
Konstantin Avrachenkov
\thanks{INRIA Sophia Antipolis, France
\texttt{K.Avrachenkov@sophia.inria.fr} }
\and
Paulo Gon\c{c}alves
\thanks{INRIA Rhone-Alpes, France
\texttt{paulo.goncalves@inria.fr}}
\and
Alexey Mishenin
\thanks{St. Petersburg State University, Russia
\texttt{Alexey.Mishenin@gmail.com}}
\and
Marina Sokol
\thanks{INRIA Sophia Antipolis, France
\texttt{marina.sokol@inria.sophia.fr}}
}
\authorhead{Avrachenkov \& Gon\c{c}alves \& Mishenin \& Sokol}
\RRtitle{Un cadre général d'optimisation pour les méthodes d'apprentissage semi-supervisées sur graphes}
\RRetitle{Generalized Optimization Framework for Graph-based Semi-supervised Learning}
\titlehead{Graph-based Semi-supervised Learning}
\RRresume{Dans ce rapport nous proposons un schéma d'optimisation générique pour l'apprentissage semi-supervisé sur des graphes.  Ce cadre intègre comme cas particuliers les approches dites du Laplacien standard et du Laplacien normalisé ainsi qu'une méthode basée sur PageRank.  Nous proposons également une interprétation probabiliste originale qui s'appuie sur la notion de marche aléatoire, puis nous étudions les comportements limites de ces méthodes. Le recours aux marches aléatoires nous permet d'expliquer les différences de performances existant entre ces trois noyaux de lissage. Une des conclusions principales de ce travail est que les méthodes construites sur PageRank sont plus robustes face au choix du paramètre de régularisation et des points marqués.  Nous illustrons nos résultats théoriques avec deux jeux de données réelles représentatives de deux défis distincts: celui des réseaux sociaux avec le cas des personnages du roman "Les Misérables" et celui des graphes d'hyper-liens à travers l'application Wikipedia. En particulier, nous démontrons qu'il est possible de classifier les articles de Wikipedia avec une très bonne précision et un très bon rappel, à partir de la seule information fournie par les liens hyper-texte.
}
\RRabstract{
We develop a generalized optimization framework for graph-based semi-supervised
learning. The framework gives as particular cases the Standard Laplacian, Normalized
Laplacian and PageRank based methods. We have also provided new probabilistic
interpretation based on random walks and characterized the limiting
behaviour of the methods. The random walk based interpretation allows us to explain
differences between the performances of methods with different smoothing kernels.
It appears that the PageRank based method is robust
with respect to the choice of the regularization parameter and the labelled data.
We illustrate our
theoretical results with two realistic datasets, characterizing different challenges:
Les Miserables characters social network and Wikipedia hyper-link graph. The graph-based
semi-supervised learning classifies the Wikipedia articles with very good precision
and perfect recall employing only the information about the hyper-text links.
}
\RRmotcle{Apprentissage Semi-supervisé, PageRank, Marche Aléatoire sur des Graphes, 
Classification Automatique des Articles de Wikipedia}
\RRkeyword{Semi-supervised Learning, PageRank, Random Walk on Graphs, Wikipedia  Automatic Article Classification}
\RRprojets{Maestro}
\RRdomaine{1} 
\RRthemeProj{maestro} 
\RRdomaineProjBis{maestro} 
\RCSophia 

\begin{document}
\RRNo{7774}
\makeRR   

\section{Introduction}

Semi-supervised classification is a special form of classification. Traditional classifiers use only
labeled data to train. Semi-supervised learning use large amount of unlabeled data, together with labeled data,
to build better classifiers. Semi-supervised learning requires less human effort and gives high accuracy.
Graph-based semi-supervised methods define a graph where the nodes are labeled and unlabeled instances in
the dataset, and edges (may be weighted) reflect the similarity of instances. These methods usually assume
label smoothness over the graph (see the excellent book on the graph-based semi-supervised learning \cite{Zhu09}).
In this work we often omit ``graph-based'' term as it is clear that we only consider graph-based
semi-supervised learning methods.

Up to the present, most literature on the graph-based semi-supervised learning studied the following two
methods: the Standard Laplacian based method (see e.g., \cite{Zhou:2007}) and the Normalized Laplacian
based method (see e.g., \cite{Zhou04learningwith}). Here we propose a generalized optimization framework which
implies the above two methods as particular cases. Moreover, our generalized optimization framework gives
PageRank based method as another particular case. The PageRank based methods have been proposed in \cite{SIGIR08}
as a classification stage in a clustering method for large hyper-text document collections. In \cite{SIGIR08}
only a linear algebraic formulation was proposed but not the optimization formulation. A great advantage
of the PageRank based method is that it has a quasi-linear complexity.
In \cite{ChungTsiatas10} a method also based on PageRank has been proposed.
However, the method of \cite{ChungTsiatas10} cannot be scaled to large datasets as it is based on the K-means method.
The generalized
optimization framework allows us to provide intuitive interpretation of the differences between particular cases.
Using the terminology of random walks on graphs we also provide new probabilistic interpretation for the Standard
Laplacian based method and the PageRank based method. With the help of the random walk terminology we are able
to explain differences in classifications provided by the Standard Laplacian based method and the PageRank based
method. The generalized optimization framework has only two free parameters to tune. By choosing the first parameter,
we vary the level of credit that we give to nodes with large degree. By choosing the second parameter, the regularization parameter, we choose a trade-off between the closeness of the classification function to the labeling function and
the smoothness of the classification function over the graph. We study sensitivity of the methods with respect
to the value of the regularization parameter. We conclude that only the PageRank based method shows robustness
with respect to the choice of the value of the regularization parameter. We illustrate our theoretical results
and obtain further insights from two datasets. The first dataset is a graph of co-appearance of the characters
in the novel Les Miserables. The second data set is a collection of articles from Wikipedia for which
we have expert classification. We have compared the quality of classification of the graph-based
semi-supervised learning methods with the quality of classification based on Wikipedia categories.
It is remarkable to observe that with just few labeled points and only using the hyper-text links,
the graph-based semi-supervised methods perform nearly as good as Wikipedia categories in terms of precision
and even better in terms of recall. With the help of the two datasets we confirm that the PageRank
based method is more robust than the other two methods with respect to the value of the regularization
parameter and with respect to the choice of labeled points.

The rest of the paper is organized as follows: In Section~2 we describe a generalized optimization
framework for the graph-based semi-supervised learning and discuss differences among particular cases.
In Section~3 we demonstrate our theoretical results by numerical examples. We conclude the paper in
Section~4 with directions for future research.

\section{Generalized Optimization Framework}

The input to a semi-supervised classification consists of a set of data instances
\(X = \{X_1,.., X_P, X_{P+1},..,X_N\} \). An instance could be described by a fixed collection
of attributes. For example, all attributes can take real numbers as values and these numbers
can be normalized. Suppose we have $K$ classes and the first $P$ instances in our dataset are
labeled as \(k(i) \in {1,...,K}\), $i=1,...,P$. Let matrix $W$ represent degrees of similarity
between instances in \(X\). The construction of $W$ can be done by various method.
If we continue with the example where attributes are given by normalized real numbers,
the Radial Basis Function (RBF)
$$
W_{ij} = \exp(-||X_i - X_j||^2 / \gamma)
$$
or k-Nearest Neighbors (kNN) method
$$
W_{ij}=
   \begin{cases}
        1, & \text{if $X_j$ is one of the k nearest neighbors of $X_i$,}\\
        0, & \text{otherwise}
   \end{cases}
$$
can be chosen to construct the similarity matrix.
In the datasets of this article we assume that the matrix \(W\) is symmetric. The RBF
method in fact gives a symmetric similarity matrix. In general, the kNN method can give
non-symmetric matrix, but it could be easily transformed to the symmetric one by $W' = (W+W^T)/2$.
Denote by $D$ a diagonal matrix with its \((i,i)\)-element equals to the sum
of the \(i\)-th row of matrix \(W\):
$$
d_{i,i}=\sum_{j=1}^{N}w_{i,j}.
$$
In some applications, which is also the case for our datasets, the similarity graph is available
as a part of the data.

Define \(N\times K\) matrix \(Y\) as\\
\begin{equation}
Y_{ik}=
   \begin{cases}
        1, & \text{if $X_i$ is labeled as $k(i)=k$,}\\
        0, & \text{otherwise.}
   \end{cases} \nonumber
\end{equation}
We refer to each column $Y_{\cdot k}$ of matrix $Y$ as labeling function.
Also define \(N\times K\) matrix \(F\) and call its columns $F_{\cdot k}$
classification functions. A general idea of the graph-based semi-supervised learning
is to find classification functions so that on the one hand they will be close
to the corresponding labeling function and on the other hand they will change
smoothly over the graph associated with the similarity matrix. This general
idea can be expressed with the help of optimization formulation. In particular,
there are two widely used optimization frameworks. The first formulation,
the Standard Laplacian based formulation \cite{Zhou:2007}, is as follows:
\begin{equation}
\label{eq:StandLaplace}
\min_{F} \{ \sum_{i=1}^N \sum_{j = 1}^N w_{ij}\|F_{i.} - F_{j.}\|^2
+ \mu\sum_{i=1}^N d_i\| F_{i.} - Y_{i.}\|^2 \}
\end{equation}
and the second, the Normalized Laplacian based formulation \cite{Zhou04learningwith},
is as follows:
\begin{equation}
\label{eq:NormLaplace}
\min_{F} \{\sum_{i=1}^N \sum_{j = 1}^N w_{ij}\|\frac {F_{i.}} {\sqrt{d_{ii}}} - \frac {F_{j.}}{\sqrt{d_{jj}}}\|^2 + \mu\sum_{i=1}^N \| F_{i.} - Y_{i.}\|^2\}
\end{equation}
where $\mu$ is a regularization parameter. In fact, the parameter $\mu$ represents
a trade-off between the closeness of the classification function to the labeling
function and its smoothness. 

Here we propose a generalized optimization framework, which has as particular cases the two above mentioned formulations. Namely, we suggest the following optimization formulation
\begin{equation}
\label{eq:general}
\min_{F} \{\sum_{i=1}^N \sum_{j = 1}^N w_{ij}\| {d_{ii}}^{\sigma-1} F_{i.}  - {d_{jj}}^{\sigma-1}F_{j.}\|^2
+ \mu\sum_{i=1}^N {d_{ii}}^{2\sigma-1} \| F_{i.} - Y_{i.}\|^2 \}
\end{equation}
In addition to the Standard Laplacian formulation ($\sigma = 1$) and the
Normalized Laplacian formulation ($\sigma = 1/2$), we obtain the third
very interesting case when $\sigma = 0$. We show below that this particular
case corresponds to PageRank based clustering \cite{SIGIR08}, for which \eqref{eq:general}
can be rewritten as:

\begin{eqnarray}
\mathop{\mbox{min}}_{F} {\sum_{i=1}^N \sum_{j = 1}^N w_{ij}\|\frac {F_{i.}} {d_{ii}} - \frac {F_{j.}}{d_{jj}}\|^2 + \mu\sum_{i=1}^N \frac {1}
{d_{ii}}\| F_{i.} - Y_{i.}\|^2} \nonumber
\end{eqnarray}

Since the objective function of the generalized optimization framework is
a sum of a positive semi-definite quadratic form and a positive quadratic form,
we can state the following proposition.

\begin{proposition}
The objective of the generalized optimization framework for semi-supervised
learning is a convex function.
\end{proposition}

One way to find $F$ is to apply one of many efficient optimization methods
for convex optimization. Another way to find $F$ is to find it as a solution
of the first order optimality condition. Fortunately, we can even find $F$
in explicit form.

\begin{proposition}
The classification functions for the generalized semi-supervised learning
are given by
\begin{equation}
\label{eq:Fgener}
F_{.k} = \frac{\mu}{2+\mu}
\left(I - \frac {2}{2+\mu}D^{-\sigma} W D^{\sigma-1}\right)^{-1} Y_{.k},
\end{equation}
for $k=1,...,K$.
\end{proposition}

{\bf Proof:}
The objective function of the generalized semi-supervised learning framework
can be rewritten in the following matrix form
\begin{eqnarray}
\begin{split}
Q(F) = 2\sum_{k=1}^K F_{.k}^TD^{\sigma-1}LD^{\sigma-1}F_{.k} \nonumber \\
+ \mu\sum_{k=1}^K (F_{.k}- Y_{.k})^TD^{2\sigma-1}(F_{.k}- Y_{.k}), \nonumber
\end{split}
\end{eqnarray}
where $L=D-W$ is the Standard Laplacian.
The first order optimality condition $D_{F_{.k}} Q(F)=0$ gives
\begin{eqnarray}
\begin{split}
2F_{.k}^T(D^{\sigma-1}LD^{\sigma-1}+D^{\sigma-1}L^TD^{\sigma-1}) \nonumber \\
+ 2\mu(F_{.k}-Y_{.k})^TD^{2\sigma-1} = 0. \nonumber
\end{split}
\end{eqnarray}
Multiplying the above expression from the right hand side by
$D^{-2\sigma+1}$, we obtain
$$
2F_{.k}^T(D^{\sigma-1}(L+L^T)D^{-\sigma}) + 2\mu(F_{.k}-Y_{.k})^T = 0.
$$
Then, substituting $L=D-W$ and rearranging the terms yields
\begin{eqnarray}
F_{.k}^T(2I - D^{\sigma-1}(W+W^T)D^{-\sigma} + \mu I ) - \mu Y_{.k}^T = 0. \nonumber
\end{eqnarray}
Since $W$ is a symmetric matrix, we obtain
\begin{eqnarray}
F_{.k}^T(2I - 2 D^{\sigma-1} W D^{-\sigma} + \mu I ) - \mu Y_{.k}^T = 0. \nonumber
\end{eqnarray}
Thus, we have
\begin{eqnarray}
F_{.k}^T = \mu Y_{.k}^T (2I - 2 D^{\sigma-1} W D^{-\sigma} + \mu I )^{-1},  \nonumber
\end{eqnarray}
which proves the proposition.

As a corollary, we have explicit expressions for the classification functions
for the three mentioned above particular semi-supervised learning methods.
Namely, from expression (\ref{eq:Fgener}) we derive
\begin{itemize}
    \item if $\sigma=1$, the Standard Laplacian method:\\
    \(F_{.k} = \frac {\mu}{2+\mu}  (I - \frac {2}{2+\mu}D^{-1}W)^{-1} Y_{.k},\)

    \item if $\sigma=1/2$, the Normalized Laplacian method:\\
    \(F_{.k} = \frac {\mu}{2+\mu}  (I - \frac {2}{2+\mu}D^{\frac{-1}{2}}WD^{\frac{-1}{2}})^{-1} Y_{.k},\)

    \item if $\sigma=0$, PageRank based method:\\
    \(F_{.k} = \frac {\mu}{2+\mu}  (I - \frac {2}{2+\mu}WD^{-1})^{-1} Y_{.k}.\)
\end{itemize}

Let us now explain why the case $\sigma=0$ corresponds to the PageRank based
clustering method. Denote $\alpha=2/(2+\mu)$ and write $F_{.k}$
in a transposed form
$$
F_{.k}^T = (1-\alpha) Y_{.k}^T (I - \alpha D^{-1}W)^{-1}.
$$
If the labeling functions are normalized, this is exactly an explicit
expression for PageRank \cite{Moler2004,Langville2006}. This expression was used in \cite{SIGIR08}
but no optimization framework was provided.

Note that $D^{-1}W$ represents the transition probability matrix for the
random walk on the similarity graph. Then, the $(i,j)$-th element of the
matrix $(I - \alpha D^{-1}W)^{-1}$ gives the expected number of visits to
node $j$ starting from node $i$ until the random walk restarts with probability $1-\alpha$.
This observation provides the following probabilistic interpretation for
the Standard Laplacian and PageRank based methods. In the Standard Laplacian
method, $F_{ik}$ gives up to a multiplicative constant the expected
number of visits before restart to the labeled nodes of class $k$ if the random walk starts
from node $i$. In the PageRank based method with normalized labeling functions,
$F_{ik}$ gives up to a multiplicative constant the expected number of visits to
node $i$, if the random walk starts from a uniform distribution over the labeled
nodes of class $k$.

The random walk approach can explain why in some cases Standard Laplacian and
PageRank based methods provide different classifications. For instance,
consider a case when a node $v$ is directly connected to the labeled nodes $k_1$ and $k_2$
belonging to different classes. Furthermore, let the labeled node $k_1$ have a higher degree
than the node $k_2$ and let the node $k_1$ belong to a denser cluster than node $k_2$.
From \cite{Avrachenkov06} we know that the expected number of visits to node $j$ starting
from node $i$ until the restart is equal to the product of the probability to visit
node $j$ before the absorption and the expected number of returns to node $j$
starting from node $j$. Then, the PageRank based method will classify the node $v$ into
the class of the labeled node $k_2$ as it is more likely that the random walk misses
the node $v$ starting from node $k_1$. In other words, when the random walk starts
from $k_2$, there are less options how to choose a next node and it is more likely
to choose node $v$ as a next node. In the Standard Laplacian method we need to compare
the average number of visits to the labeled nodes starting from the node $v$. Since
the random walk can reach either node $k_1$ or node $k_2$ in one step the probabilities
of hitting these nodes before absorption are similar and what matters is how dense are
the classes. If the class associated with the labeled node $k_1$ is more dense than
the class associated with the labeled node $k_2$, the node $v$ will be classified to
the class associated with $k_1$. We shall illustrate the above reasoning by a specific
example in the next section.
\begin{figure}[h!]
\begin{center}
    \subfigure[Smoothness term]
    {
        \includegraphics[height=2in]{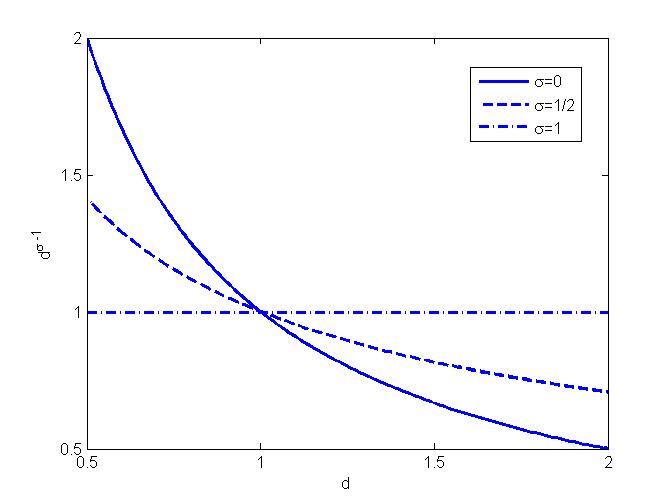}
        \label{fig:Smooth}
    }
    \subfigure[Fitting term]
    {
        \includegraphics[height=2in]{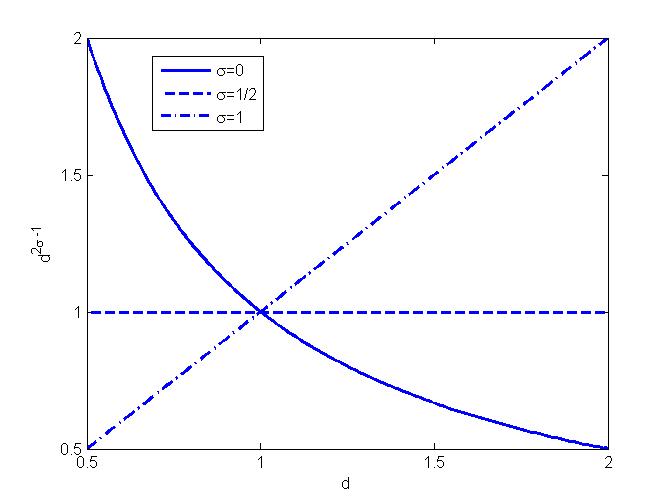}
       \label{fig:Fitting}
    }
\end{center}
\caption{Fitting and smoothness terms}
\end{figure}

Based on the formulation~\eqref{eq:general}, we could give some further
intuitive interpretation for various cases of the generalized semi-supervised
learning.
Let us consider the first term in the r.h.s. sum of \eqref{eq:general}, which corresponds
to the smoothness component.  Figure~\ref{fig:Smooth} shows that if $\sigma < 1$ we do not
give much credit to the connections between points with large degrees.
Let us now consider the second term which corresponds to the fitting function. Figure~\ref{fig:Fitting}
shows that $\sigma<1/2$ does not give much credit to samples that pertain to a dense cluster of points (i.e. $d_{ii}$ is large), whereas samples that are relatively isolated in the feature space (corresponding to small
value of $d_{ii}$), are given higher confidence. If $\sigma = 1$, the node degree does not have any influence.
And if $\sigma > 1/2$, we consider that the nodes with higher weighted degree are more important
than the nodes with smaller degree.
%
%

Next we analyze the limiting behavior of the semi-supervised learning
methods when $\mu \to 0$ ($\alpha \to 1$). We shall use the following
Laurent series expansion
\begin{equation}
\label{eq:Laurent}
(1-\alpha)[I-\alpha D^{-1}W]^{-1} = \alpha [\ones \pi + (1-\alpha)H +\mbox{o}(1-\alpha)],
\end{equation}
where $\pi$ is the stationary distribution of the random walk
($\pi D^{-1}W = \pi$), $\ones$ is a vector of ones of appropriate
dimension and $H=(I-D^{-1}W+\ones \pi)^{-1}-\ones \pi$ is the
deviation matrix \cite{Puterman1994}. Let us note that if the similarity matrix $W$
is symmetric, the random walk governed by the transition matrix $D^{-1}W$ is time-reversible
and its stationary distribution is given by
\begin{equation}
\label{eq:revdistr}
\pi = (\ones^T D \ones)^{-1} \ones^T D.
\end{equation}

Let us insert the Laurent series expansion (\ref{eq:Laurent}) into the expression for
the general classification function (\ref{eq:Fgener}):
\begin{equation}
\label{eq:LimitGener}
\begin{split}
F_{.k} &= (1-\alpha) (I - \alpha D^{-\sigma} W D^{\sigma-1})^{-1} Y_{.k}  \\
&= (1-\alpha) [D^{-\sigma+1}(I - \alpha D^{-1} W) D^{\sigma-1}]^{-1} Y_{.k} \\
&= (1-\alpha) D^{-\sigma+1}[I - \alpha D^{-1} W]^{-1} D^{\sigma-1} Y_{.k} \\
&= \alpha D^{-\sigma+1}[\ones \pi + (1-\alpha)H +\mbox{o}(1-\alpha)] D^{\sigma-1} Y_{.k} \\
&= \alpha [D^{-\sigma+1}\ones \sum_{i : k(i)=k} \pi_i Y_{ik} d^{\sigma-1}_i  \\
&+ (1-\alpha)D^{-\sigma+1}HD^{\sigma-1}Y_{.k} +\mbox{o}(1-\alpha)].
\end{split}
\end{equation}
Next, using the expression for the stationary distribution (\ref{eq:revdistr}), we can specify (\ref{eq:LimitGener})
as follows:
\begin{equation}
\begin{split}
F_{.k} &= \alpha [D^{-\sigma+1}\ones (\ones^T D \ones)^{-1} \sum_{i : k(i)=k} Y_{ik} d^{\sigma}_i  \\
&+ (1-\alpha)D^{-\sigma+1}HD^{\sigma-1}Y_{.k} +\mbox{o}(1-\alpha)].
\end{split}
\end{equation}
Hence, in the Semi-supervised methods when $\alpha$ is sufficiently
close to one, a class with the largest $\sum_{i : k(i)=k} Y_{ik} d^{\sigma}_i$
attracts all instances.
This implies that the limiting behavior of the PageRank based method ($\sigma=0$)
is quite different from the limiting behaviors of the other
methods. In particular, if the number of labelled points in each class is
same or if the labeling functions are normalized, then there is no dominating
class which attracts all instances and the classification results most likely
make sense even for $\alpha$ very close to one. The conclusion is that the
PageRank based method is more robust to the choice of the regularization parameter
than the other graph-based semi-supervised learning methods.

\noindent
{\bf Illustrating example:} to illustrate the limiting behaviour of the methods we generated an artificial example of the planted partition random graph model \cite{Condon2001} with two classes with 100 nodes in each class.
The probability of link creation inside the first class is 0.3 and
the probability of link creation inside the second class is 0.1.
So the first class is three times denser than the second class.
The probability  of link creation between two classes is 0.05.
We have generated a sample of this random graph model.
In each class we have chosen just one labelled point. In the first class we have chosen as the labelled point
the point with the smallest degree (degree=28, 24 edges inside the class and 4 edges leading outside).
In the second class we have chosen as the labelled point the point with the largest degree
(degree=31, 27 edges inside the class and 4 edges leading outside).
We have indeed observed that the second class attracts all points when $\alpha$ is close to one for all semi-supervised methods except for the PageRank based method. This is in accordance with theoretical conclusions as the labelled point in the second class has a larger weight than the labelled point in the first class. It is interesting to observe that in this example the first class looses all points when $\alpha$ is close to one even though the first class is denser then the second one.
\newpage

\section{Experiments}

In this section we apply the developed theory to two datasets. The first dataset is the network of interactions
between major characters in the novel Les Miserables. If two characters participate in one or more scenes, there is
a link between these two characters. The second dataset is a subset of Wikipedia pages. Wikipedia articles correspond
to the data points and hyper-text links correspond to the edges of the similarity graph. We disregard the
direction of the hyper-text links.


\subsection{Les Miserables example}

\begin{figure}[h]
\begin{center}
    \subfigure[Modularity as a function of $\alpha$.]
    {
        \includegraphics[width=3.5in]{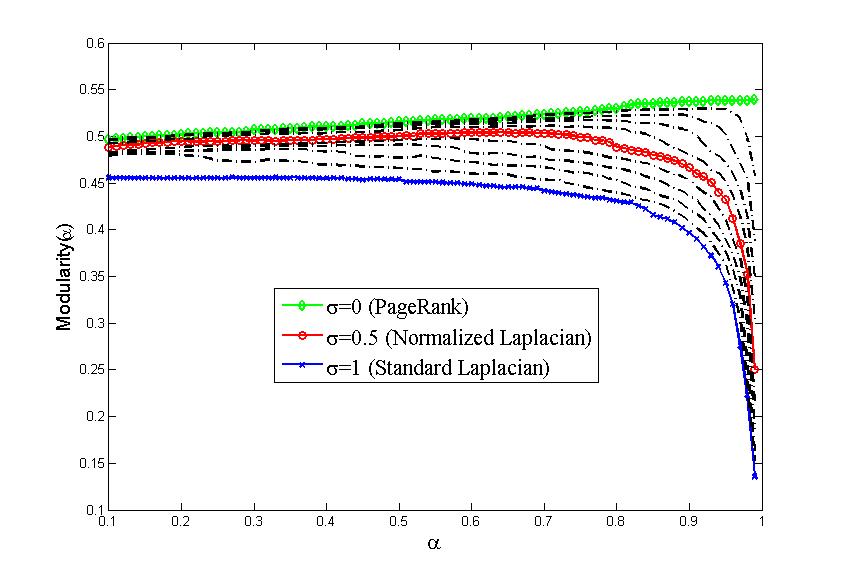}
        \label{fig:LesMisRandSeedsMod}

    }
    \subfigure[Difference in classifications.]
    {
       \includegraphics[width=3.3in]{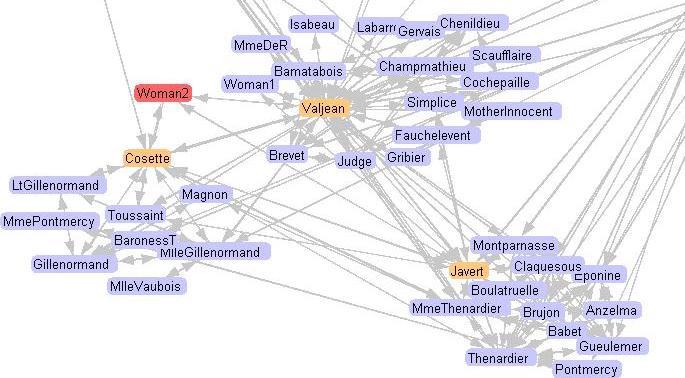}
       \label{fig:LesMis4}

    }
\end{center}
\caption{Les Miserables example.}
\end{figure}

The graph of the interactions of Les Miserables characters has been compiled by Knuth \cite{Knuth1993}.
There are 77 nodes and 508 edges in the graph. Using the betweenness based algorithm
of Newman \cite{Newman2004} we obtain 6 clusters which can be identified with the main characters:
Valjean (17), Myriel (10), Gavroche (18), Cosette (10), Thenardier (12), Fantine (10),
where in brackets we give the number of nodes in the respective cluster.
We have generated randomly 100 times labeled points (one labeled point per cluster).
In Figure~\ref{fig:LesMisRandSeedsMod} we plot the modularity measure averaged over 100
experiments as a function of $\alpha$  for methods with different values of $\sigma$ ranging from 0 to 1 with granularity 0.1.
The modularity measure is based on the inter-cluster link density and the average link density
and reflects the quality of clustering \cite{Newman2004}.
From Figure~\ref{fig:LesMisRandSeedsMod} we conclude that on average the PageRank based method performs best
in terms of modularity and it is robust with respect to the choice of the regularization parameter.
In particular, we observe that as was predicted by the theory the Standard Laplacian method
and Normalized Laplacian method perform badly when $\alpha$ is close to 1 (one class attracts
all instances). The PageRank based method is robust even for the values of $\alpha$ which are very
close to one.

Next let us use the random walk based interpretation to explain differences between the Standard Laplacian
based method and the PageRank based method. Let us consider the node Woman~2 (see Figure~\ref{fig:LesMis4}).
The node Woman~2 is connected with three other nodes: Valjean, Cosette and Javert.
Suppose we have chosen labeled points so that only the nodes Valjean and Cosette are labeled but not
Javert. Since the node Valjean has many more links than the node Cosette, the random walk starting from the
node Valjean will less likely hit the node Woman~2 than the random walk starting from the node Cosette in some
given time. Thus, the PageRank based method classifies the node Woman~2 into the class corresponding to Cosette.
Since the node Woman~2 is just one link away from both Valjean and Cosette, the probability to hit these nodes
before absorption is approximately equal. Thus, if we apply the Standard Laplacian method the classification
will be determined by the expected number of returns to the labeled nodes before absorption. Since the labeled node
Valjean lies in the larger and denser class, the Standard Laplacian method classifies the node
Woman~2 into the class corresponding to Valjean.

\subsection{Wikipedia-math example}

\begin{figure}[h!]
\begin{center}
    \subfigure[Modularity as a function of $\alpha$.]
    {
         \includegraphics[width=3.5in]{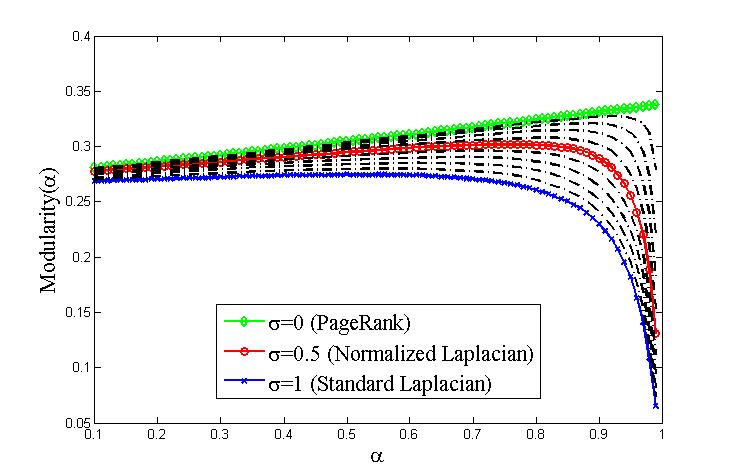}
         \label{fig:WikiMathRandSeedsMod}

    }
    \subfigure[Precision as a function of $\alpha$.]
    {
       \includegraphics[width=3.5in]{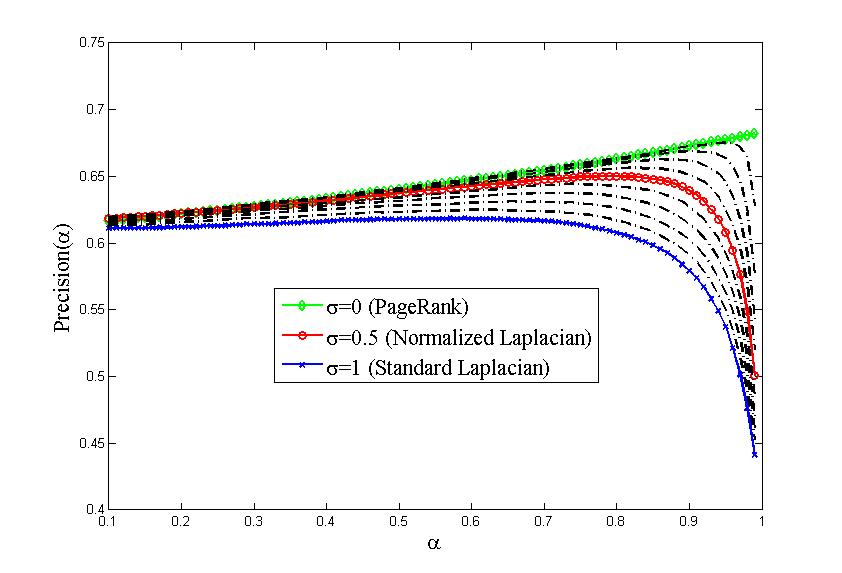}
       \label{fig:WikiMathRandSeedsPrec}
    }
\end{center}
\caption{Wikipedia-math example}
\end{figure}

The second dataset is derived from the English language Wikipedia. In this case, the similarity graph is constructed
by a slight modification of the hyper-text graph. Each Wikipedia article typically contains links to other Wikipedia articles which are used to explain specific terms and concepts. Thus, Wikipedia forms a graph whose nodes represent articles and whose edges represent hyper-text inter-article links. For our experiments we took a snapshot (dump) of Wikipedia from January 30, 2010\footnote{\textbf{\texttt{http://download.wikimedia.org/enwiki/20100130}}}. Based on
this dump we have extracted outgoing links for other articles. The links to special pages (categories, portals, etc.) have been ignored. In the present experiment we did not use the information about the direction of links, so the graph in our experiments is undirected. Then we have built a subgraph with mathematics related articles, a list of which was obtained from ``List of mathematics articles'' page from the same dump. In the present experiments we have chosen
the following three mathematical topics: ``Discrete mathematics'' (DM), ``Mathematical analysis'' (MA), ``Applied mathematics'' (AM). With the help of AMS MSC Classification \footnote{\textbf{\texttt{http://www.ams.org/mathscinet/msc/msc2010.html}}} and experts we have classified related Wikipedia mathematical articles into the
three above mentioned topics. According to the expert annotation we have built a subgraph of the Wikipedia mathematical articles providing imbalanced classes DM (106), MA (368) and AM (435). The subgraph induced by these three topics
is connected and contains 909 articles. Then, the similarity matrix $W$ is just the adjacency matrix of this subgraph. Thus, $w_{ij}=1$ means that Wikipedia article $i$ is connected with Wikipedia article $j$. Then, we have chosen uniformly
at random 100 times 5 labeled nodes for each class. In Figure~\ref{fig:WikiMathRandSeedsMod} we plot the modularity
averaged over 100 experiments as a function of $\alpha$ for methods with different values of $\sigma$ ranging from 0 to 1
with granularity 0.1.
Figure~\ref{fig:WikiMathRandSeedsMod} confirms the observations obtained
from Les Miserable dataset that the PageRank based method ($\sigma=0$) has the best performance in terms of the modularity measure.
Next, in Figure~\ref{fig:WikiMathRandSeedsPrec} we plot the precision as a function of the regularization
parameter for each of the three methods with respect to the expert classification. For the most values of $\alpha$
the PageRank based method performs better than all the other methods and shows robust behaviour when the
regularization parameter approaches one. This is in agreement with the theoretical conclusions at the
end of Section~2. Both Figure~\ref{fig:WikiMathRandSeedsMod} and Figure~\ref{fig:WikiMathRandSeedsPrec}
demonstrate that the PageRank based method is also more robust than the other two methods with respect to
the choice of labeled points.

\begin{figure}[h!]
\begin{center}
    \subfigure[Modularity as a function of $\alpha$ for PageRank.]
    {
         \includegraphics[width=3.5in]{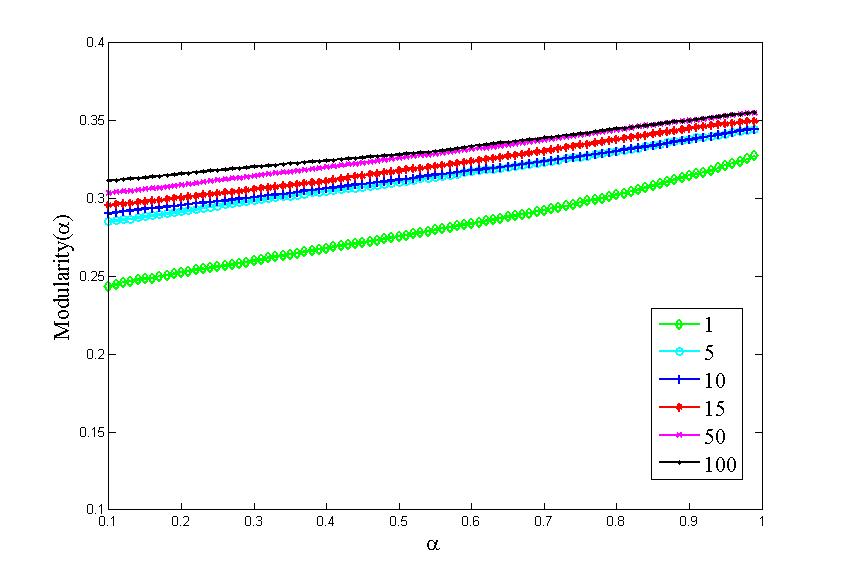}
         \label{fig:WikiMathRandSeedsModLabsPR}

    }
    \subfigure[Modularity as a function of $\alpha$ for Normalized Laplacian.]
    {
       \includegraphics[width=3.5in]{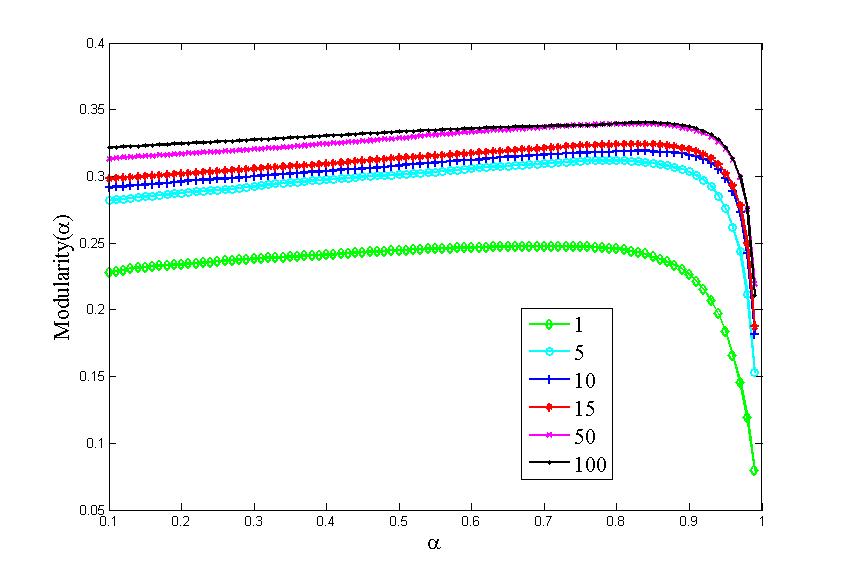}
       \label{fig:WikiMathRandSeedsModLabsNL}
    }
    \subfigure[Modularity as a function of $\alpha$ for Standard Laplacian.]
    {
       \includegraphics[width=3.5in]{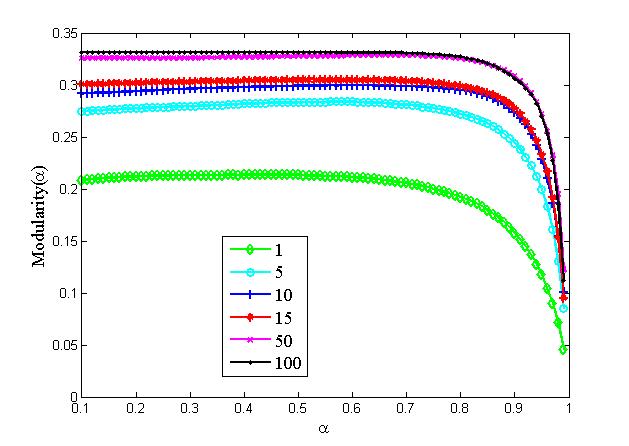}
       \label{fig:WikiMathRandSeedsModLabsSL}
    }
\end{center}
\caption{Wikipedia-math example}
\end{figure}

\begin{figure}[h!]
\begin{center}
    \subfigure[Precision as a function of $\alpha$ for PageRank.]
    {
         \includegraphics[width=3.6in]{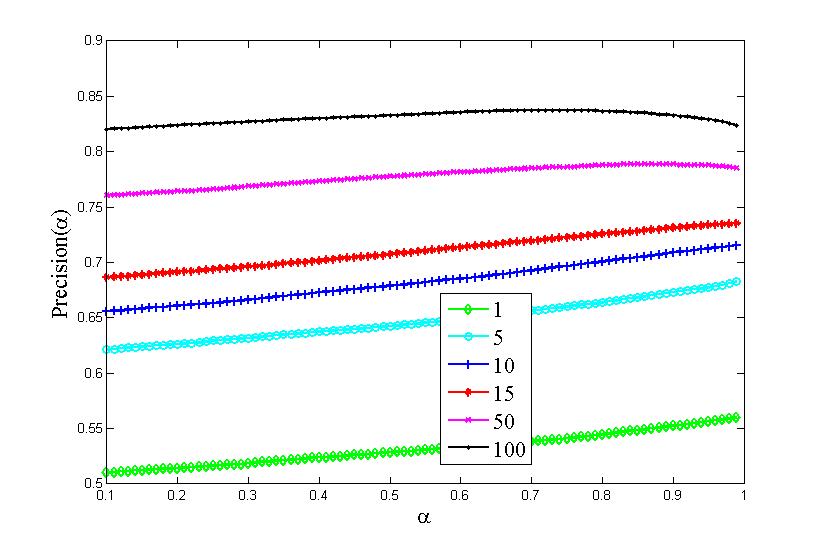}
         \label{fig:WikiMathRandSeedsPrecLabsPR}

    }
    \subfigure[Precision as a function of $\alpha$ for Normalized Laplacian.]
    {
       \includegraphics[width=3.6in]{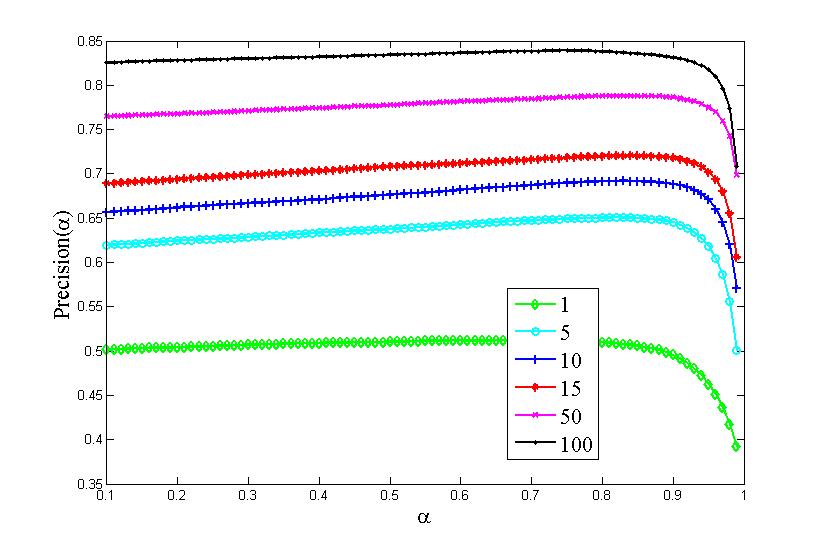}
       \label{fig:WikiMathRandSeedsPrecLabsNL}
    }
    \subfigure[Precision as a function of $\alpha$ for Standard Laplacian.]
    {
       \includegraphics[width=3.6in]{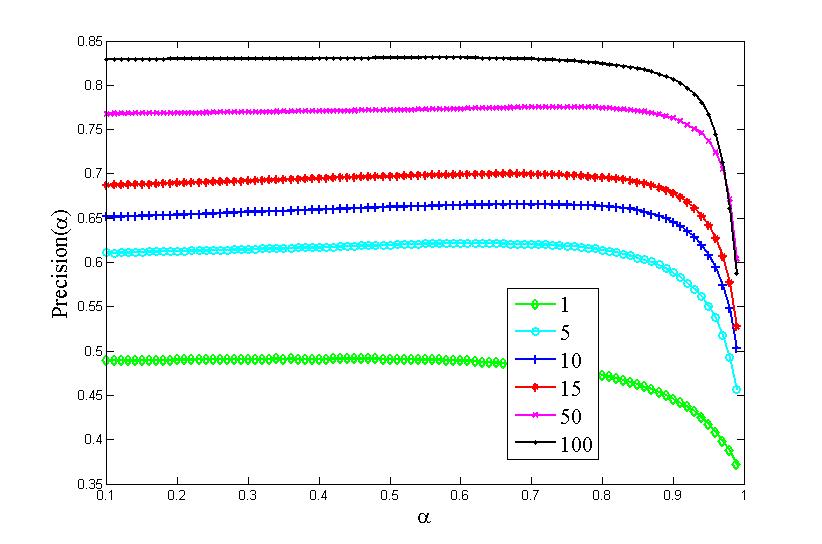}
       \label{fig:WikiMathRandSeedsPrecLabsSL}
    }
\end{center}
\caption{Wikipedia-math example}
\end{figure}

Figure~\ref{fig:WikiMathRandSeedsMod} and Figure~\ref{fig:WikiMathRandSeedsPrec} also suggest that we can use
the modularity measure as a good criterion for the choice of the regularization parameter for the Standard Laplacian and
Normalized Laplacian methods.
Now, let us investigate the effect of the quantity of the labelled data on the quality of the classification.
Figures~\ref{fig:WikiMathRandSeedsModLabsPR}~\ref{fig:WikiMathRandSeedsModLabsNL}~and~\ref{fig:WikiMathRandSeedsModLabsSL}
show that on average the modularity of the classification increases when we increase the quantity of the labelled data.
Moreover, the quality of classification improves significantly when we increase the quantity of labelled data for each class from few points to about 50 points. The further increase of the quantity of the labelled data does not result in significant improvement in classification quality. The same behaviour manifests itself with respect to the precision measure (see Figures~\ref{fig:WikiMathRandSeedsPrecLabsPR}~\ref{fig:WikiMathRandSeedsPrecLabsNL}~and~\ref{fig:WikiMathRandSeedsPrecLabsSL}).

Both Les Miserables and Wikipedia-math datasets indicate that for the PageRank based
method it is better to choose the value of the regularization parameter as close to one as possible
but at the same time keeping the system numerically stable and efficient. This is an example of the
singular perturbation phenomena \cite{Avrachenkov99,Yin1998}



We have also compared the results obtained by the semi-supervised learning methods with the classification provided
by Wikipedia Categories. As Wikipedia categories we have chosen: {\tt Applied\_mathematics, Mathematical\_analysis}
and {\tt Discrete\_mathematics}. It turns out that the precision of the Wikipedia categories with respect to the
expert classification is 78\% (with 5 random labelled points the PageRank based method can achieve about 68\%).
However, the recall of the Wikipedia categorization is 72\%. With the help of the semi-supervised learning approach
we have classified all articles. It is quite interesting to observe that just using the link information the semi-supervised
learning can achieve precision nearly as good as the Wikipedia categorization produced by hard work of many experts
and the semi-supervised learning can do even better in terms of recall.
\newpage

\section{Conclusion and future research}

We have developed a generalized optimization approach for the graph-based semi-supervised
learning which implies as particular cases the Standard Laplacian, Normalized
Laplacian and PageRank based methods and provides the new ones based on parameter $\sigma$. We have provided new probabilistic
interpretation based on random walks. This interpretation allows us to explain
differences in the performances of the methods. We have also characterized the limiting
behaviour of the methods as $\alpha \to 1$ which based on the weight of the labelled points.
We have illustrated theoretical results with the help of Les Miserables example and
Wikipedia-math example. Also, we show how the number of labeled points has an influence on the quality of the classification.
Both theoretical and experimental results demonstrate that
the PageRank based method outperforms the other methods in terms of clustering modularity
and robustness with respect to the choice of labelled points and regularization
parameter. We propose to use the modularity measure
for the choice of the regularization parameter in the cases of the Standard Laplacian
method and the Normalized Laplacian method. In the case of the Pagerank based method
we suggest to choose the value of the regularization parameter as close to one as
possible but at the same time keeping the system numerically stable and efficient.
It appears that remarkably we can classify the Wikipedia articles
with very good precision and perfect recall employing only the information about
the hyper-text links. As future research we plan to apply the cross-validation
technique to the choice of the kernel and to apply our approach to inductive
semi-supervised learning \cite{Altun05, Guo08}, which will help us to work with
out-of-sample data.

\newpage

\section*{Acknowledgement}

The work has been supported by the joint INRIA Alcatel-Lucent Laboratory.
\newpage

\bibliographystyle{plain}
\bibliography{NIPSArticle}
\newpage
\tableofcontents
\newpage

\end{document}